\documentclass[aps,showpacs,preprintnumbers,amsmath,amssymb,nofootinbib,epsf]{revtex4}
\usepackage{graphicx}

\begin{document}

\preprint{gr-qc/0505014}

\title{Radiation transport equations in non-Riemannian space-times}

\author{K. S. Cheng}
\email{hrspksc@hkucc.hku.hk} \affiliation{ Department of Physics,
The University of Hong Kong, Pokfulam Road, Hong Kong, Hong Kong
SAR, P. R. China}

\author{T. Harko}
\email{harko@hkucc.hku.hk} \affiliation{ Department of Physics,
The University of Hong Kong, Pokfulam Road, Hong Kong, Hong Kong
SAR, P. R. China}

\author{X. Y. Wang}
\email{xywang@nju.edu.cn} \affiliation{ Department of Astronomy,
Nanjing University, Nanjing, P. R. China}

\date{\today}

%% REVTEX4

\begin{abstract}

The transport equations for polarized radiation transfer in
non-Riemannian, Weyl-Cartan type space-times are derived, with the
effects of both torsion and non-metricity included. To obtain the
basic propagation equations we use the tangent bundle approach.
The equations describing the time evolution of the Stokes
parameters, of the photon distribution function and of the total
polarization degree can be formulated as a system of coupled first
order partial differential equations.  As an application of our
results we consider the propagation of the cosmological gamma ray
bursts in spatially homogeneous and isotropic spaces with torsion
and non-metricity. For this case the exact general solution of the
equation for the polarization degree is obtained, with the effects
of the torsion and non-metricity included. The presence of a
non-Riemannian geometrical background in which the electromagnetic
fields couple to torsion and/or non-metricity affect the
polarization of photon beams. Consequently, we suggest that the
observed polarization of prompt cosmological gamma ray bursts and
of their optical afterglows may have a propagation effect
component, due to a torsion/non-metricity induced birefringence of
the vacuum. A cosmological redshift and frequency dependence of
the polarization degree of gamma ray bursts also follows from the
model, thus providing a clear observational signature of the
torsional/non-metric effects. On the other hand, observations of
the polarization of the gamma ray bursts can impose strong
constraints on the torsion and non-metricity and discriminate
between different theoretical models.

\end{abstract}

%% REVTEX4

\maketitle

\section{Introduction}

The study of the propagation of electromagnetic radiation in
gravitational fields plays an essential role in relativistic
astrophysics and cosmology. In order to obtain a correct and
consistent description of the radiative transfer processes for
photons propagating on cosmological distances or on the cosmic
microwave background one must also take into account the effects
of the background geometry on the radiation emitted near the
surface of neutrons stars or black holes \cite{RiMe88}. The
radiation follows curved paths according to the underlying
geometry and is red or blue-shifted. Spatial curvature in some
geometries, like, for example, the Kerr geometry, rotates the
polarization of photons propagating through them.

Generally, one cannot solve Maxwell's equations exactly for waves
propagating in curved space-times or in a relativistic medium.
Instead one either uses a geometrical optics approximation or a
kinetic description of the multi-photon system, by treating
photons as massless classical particles, characterized by a
four-momentum $p$ at an event $x$ \cite{BrEh80}. In the
geometrical optics approximation, which is reasonable in most of
the cases, the degree of polarization is not affected by the
underlying geometry. The field vector propagate parallel along the
light paths. For non-flat curved paths, a rotation of the
polarization angle results.

Astrophysicists measure intensities and spectral distributions. These can
conveniently be described by a distribution function, which is defined on
phase space and is directly related to the spectral intensity. In this
picture the propagation of radiation is described by an equation of
radiative transfer, which is a differential equation for the photon
distribution function. In addition to the above-mentioned quantities,
degrees and plane of polarization of the observed radiation provide
important means of information about the source and the intervening medium
and geometry.

A general relativistic form of the Boltzmann transport equation
for particles or radiation interacting with an external medium has
been developed by Lindquist \cite{Li66}. The comoving frame
radiation transfer equation in an Eulerian coordinate system was
obtained by Riffert \cite{Ri86}. A transport equation for photons
in curved Riemannian space-times was derived by Bildhauer
\cite{Bi89}, by making use of the covariant generalization of the
Wigner transformation. Two linear equations on the tangent bundle
are obtained whose correction terms to the Liouville equation and
the classical mass-shell condition are of the order of the
wavelength over the variation length of the background geometry.
The comoving frame transfer equation for the Stokes parameters in
an arbitrary background metric in an Eulerian coordinate system
has been obtained in \cite{Bi89a} and applied to the study of the
Berry's phase for polarized light in \cite{Bi90}.

A long time ago Brans \cite{Br75} has suggested that a rotation of
the polarization plane of the radiation arises whenever there is
shear, by an effect analogous to the Thomas precession. Therefore
if there is a large scale anisotropy in the expansion of the
Universe, the microwave background radiation is expected to be
linearly polarized. In Friedmann-Robertson-Walker models with
small expansion anisotropy the observed rotation of the
polarization plane would be appreciable and constant over the
celestial sphere in closed (Bianchi type IX) models or it should
vanish in flat (type I) and open (type V) models \cite{MaTo82}.
Hence the study of the polarization of the cosmic microwave
background radiation can imposes strong limits on the anisotropy
and other fundamental physical parameters of the Universe. A
rigorous first order solution with respect to the anisotropy for
the equation of polarized radiation transfer in a homogeneous
anisotropic Universe was obtained in \cite{BaPo80}. The degree of
polarization of the background radiation is very sensitive to the
recombination dynamics and to the reheating epoch. For a Bianchi
type-II space-time with a frozen-in magnetic field the
polarization and anisotropy properties of the cosmic microwave
background radiation have been studied in \cite{FaTa87}.
Faraday-rotation effects, such as the rotation of the linear
polarization plane, are found to be independent of the spatial
curvature effect and are thus the same as in type-I models.

Almost immediately after the birth of general relativity (GR),
more general geometries, with non-metricity and torsion, have been
proposed by Weyl and Cartan, in order to incorporate in a
geometric framework the effects of the electromagnetism and of the
angular momentum. Later one, these extensions of the classical GR
have been incorporated in the different gauge theoretical
formulations of gravity \cite{He95}. There are different gauge
theories of gravity in dependence of the choice of the
gravitational gauge group and of the gravitational Lagrangian. One
of the most studied gauge theory of gravity is the Poincar\'e
gauge theory, which requires a generalization of the Riemannian
geometry and the introduction of the torsion of the space-time
\cite{He76}. Non-symmetric gravitation theories, in which the
gravitational field is described by a non-symmetric metric tensor,
whose antisymmetric part couples directly to the electromagnetic
field, have also been proposed \cite{MoWo88}.

The study of the propagation of light in a gravitational field has
provided the classic observational tests of general relativity.
The modifications of the geometry due to the non-Riemannian
effects, like torsion and non-metricity, further affect the motion
of photons, by inducing a birefringence of the vacuum which can
modify the polarization of the photons as they propagate through
the gravitational field. Thus the study of the polarization of
photons from astrophysical or cosmological sources can provide a
valuable tool for discriminating between the different
modifications and extensions of general relativity. Ni \cite{Ni84}
has shown that non-metric gravitational fields can single out
linear polarization states of light that propagate with different
speeds. He also proposed to use pulsar polarization data to impose
constraints on non-metric theories. Limits on the gravity-induced
polarization of the Zeeman components of the solar spectral lines
have been obtained in \cite{GaHaMaPa91}. The polarization of light
from magnetic white dwarfs can be used to impose constraints on
the gravity-induced birefringence of space in nonmetric
gravitational theories \cite{SoHaMa99}. By using the magnetic
white dwarf Grw+70$^{\circ }$ 8247 polarization data one can
obtain the constraint $l_*^2<(4.9\;{\rm km})^2$, where $l_*$ is
the non-symmetric gravitational theory charge.

In conventional Maxwell-Lorentz electrodynamics, the propagation
of light is influenced by the metric only, and not by the torsion
$T$ of the space-time. However, there is a possibility of
interaction between light and torsion if the latter is
non-minimally coupled to the electromagnetic field $F$ by means of
a Lagrangian of the form $\sim l^2F^2T^2$, where $l$ is a coupling
constant. Several such couplings have been proposed and analyzed
recently. A supplementary Lagrangian of the form $L=\lambda
_0l^2\left(T_\alpha \wedge F\right)T^\alpha \wedge F$ was
considered in \cite{RuObHe03}, and it was shown that it can yield
birefringence in the vacuum. Non-minimally coupled homogeneous and
isotropic torsion field in a Friedmann-Robertson-Walker geometry
affects the speed of light, with the photons propagating with a
torsion-dependent speed. In fact, torsion generates three major
effects affecting the propagation of light: it produces an axion
field that induces an optical activity into space-time, modifies
the light cone structure that yields birefringence of the vacuum
and modifies the speed of light in a torsion-dependent way
\cite{ItHe03}. An example of a metric-affine gauge theory of
gravity in which torsion couples nonminimally to the
electromagnetic field was considered in \cite{So04}. The coupling
causes a phase difference to accumulate between different
polarization states of light as they propagate through the
metric-affine gravitational field. The model has been constrained
by using solar spectropolarimetric observations, which allow to
set an upper bound on the relevant coupling constant $k$, $k^2 <
\left(2.5 {\rm km}\right)^2$.

The confirmation that at least some gamma-ray bursts (GRBs) are
indeed at cosmological distances raises the possibility that
observations of these could provide interesting constraints on the
fundamental laws of physics (for reviews on GRBs see
\cite{ChLu01}-\cite{Pi04}). The fine-scale time structure and hard
spectra of GRB emissions are very sensitive to the possible
dispersion of electromagnetic waves in vacuo. Hence the study of
short-duration photon bursts propagating over cosmological
distances is the most promising way to probe the quantum
gravitational and/or the effects related to the existence of
extra-dimensions \cite{HaCh03}. The modification of the group
velocity of the photons by the quantum effects would affect the
simultaneity of the arrival times of photons with different
energies. Thus, given a distant, transient source of photons, one
could measure the differences in the arrival times of sharp
transitions in the signals in different energy bands and use this
information to constrain quantum gravity and/or multidimensional
effects \cite{ChHa03}.

The announcement of the results of the polarization measurements
of the prompt gamma-rays and of the optical afterglows of
gamma-ray bursts attracted a great interests recently. Coburn and
Boggs \cite{CoBo03} claimed the measurement of a very large linear
polarization, $80\pm 20 \%$, of the prompt gamma-rays from an
extremely bright burst, GRB 021206. However, this polarization
measurement has been criticized by Rutledge and Fox \cite{RuFo04},
who obtained an upper limit of only $<4.1\%$ at $90\%$ confidence
from the same data, while in Wigger et al. \cite{Wi04} a value of
41\% (+57\% -44\%) is found. One the other hand, the polarization
measurements of the optical afterglows are much more convincing.
The first positive detection of the polarization was for GRB
990510 with the degree of $1.7\pm0.2\%$, and since then polarized
emission has been measured in several other afterglows (see
\cite{Co02} for a review).

It is the purpose of the present paper to generalize the transport
equations for polarized radiation \cite{Li66}, \cite{Ri86},
\cite{Bi89}, \cite{Bi89a} for the case of the polarized radiation
propagating in a space-time manifold with non-Riemannian geometry.
More exactly, we shall consider the case of the so-called
Weyl-Cartan space-times, whose geometrical structure is described
by three tensors: the metric, the torsion and the non-metricity
\cite{He76}. In this case the propagation equations for the
radiation distribution function, for the Stokes parameters, for
the linear polarization and for the degree of the total
polarization can be formulated as a system of a first order
partial differential equations on the photon phase-space, with the
non-Riemannian effects included via the contorsion tensor.

As an astrophysical application of the obtained equations we
consider the propagation of gamma ray bursts in a non-Riemannian
geometrical background. The general solution of the propagation
equation for the total polarization degree can be obtained in an
exact form. The presence of a non-Riemannian geometrical
background, in which the electromagnetic fields couple to torsion
and/or non-metricity affect the polarization of photon beams. The
polarization of the beam is generally a function of time, with
non-Riemannian effects induced via the contorsion tensor. Since
most of the gamma ray bursts have a cosmological origin, due to
the long propagation times, the influence of the torsion and
non-metricity could significantly affect their polarization.
Therefore we suggest that the observed polarization of the prompt
gamma ray bursts and of their optical afterglows could also
contain a propagation effect component, due to a
torsion/non-metricity induced birefringence of the vacuum. If such
a component could be unambiguously detected, this would provide a
significant test for the existence of the non-Riemannian
geometrical effects in our Universe. However, there are many
conventional astrophysical mechanisms that could explain the
polarization of the gamma ray bursts, and affect the propagation
of the electromagnetic radiation. The uncertainties in the
knowledge of the astrophysical environment in which photons
propagate, and the difficulty in separating the different physical
effects contributing to the polarization of the gamma ray bursts
make the practical detection of the torsion or non-metricity an
extremely difficult and very challenging observational task.

The present paper is organized as follows. The transport equations
for polarized light in Weyl-Cartan space-times are derived in
Section II. The general solution for the total polarization degree
for a homogeneous and isotropic background geometry is obtained in
Section III. Constraints on torsion and non-metricity obtained
from the observed polarization of gamma ray bursts are obtained in
Section IV. In Section V we discuss and conclude our results.

\section{Transport equations for polarized light in Weyl-Cartan
space-times}

Relativistic transport theory finds its most elegant and natural
expression in terms of geometric structures defined in the tangent
bundle over the space-time manifold.

Consider a time-oriented Lorentzian four-dimensional space-time manifold $M$%
, with metric $g$ of signature $\left( +,-,-,-\right) $. The tangent bundle $%
T(M)$ is a real vector bundle whose fibers at a point $x\in M$ is
given by the tangent space $T_{x}\left( M\right) $. If $X$ is a
tangent vector and $s$ is a section of $T(M)$, then a connection
$\nabla $ is a rule $\nabla _{X}\left( s\right) $ for taking the
directional derivative of $s$ in the direction $X$ satisfying the
properties of linearity in $s$ and $X$,
behaving like a first-order differential operator and being tensorial in $X$%
. The curvature operator is defined as $R\left( X,Y\right) =\nabla
_{x}\nabla _{Y}\left( s\right) -\nabla _{Y}\nabla _{X}\left(
s\right)
-\nabla _{\left[ X,Y\right] }\left( s\right) $, where $R\left( \frac{%
\partial }{\partial x^{\mu }},\frac{\partial }{\partial x^{\nu }}\right)
\left( {\bf e}_{i}\right) ={\bf e}_{j}R_{i\mu \nu }^{j}$, where
$\left\{{\bf e}_{1},{\bf e}_{2},...,{\bf e}_{k}\right\} $ is a
local frame defined on each neighborhood $U\subset M$. The
curvature operator satisfies the properties of multi-linearity,
anti-symmetry and tensoriality. The torsion operator on $T(M)$ is
defined as $T\left( X,Y\right) =\nabla _{X}Y-\nabla _{Y}X-\left[
X,Y\right] $. This is a vector field with components $T\left(
\frac{\partial }{\partial x^{\mu }},\frac{\partial }{\partial x^{\nu }}%
\right) =\left( \Gamma _{\mu \nu }^{\lambda }-\Gamma _{\nu \mu
}^{\lambda
}\right) \frac{\partial }{\partial x^{\lambda }}=T_{\mu \nu }^{\lambda }%
\frac{\partial }{\partial x^{\lambda }}$, where $\Gamma _{\mu \nu
}^{\lambda }$ are the Christoffel symbols on $T(M)$ and $T_{\mu
\nu }^{\lambda }$ is the torsion tensor \cite{EgGiHa80}. Moreover,
we assume that the space-time manifold $M$ is equipped with a
conformal structure, i.e. with a class $[g]$ of conformally
equivalent Lorentz metrics (and not a definite metric as in
general relativity). This corresponds to the requirement that it
should only be possible to compare lengths at one and the same
world point. We also suppose that the connection $\nabla $
respects the conformal structure. Differentially this means that
for any $g\in \left[ g\right] $ the covariant derivative $\nabla
g$ should be proportional to $g$: $\nabla g=-2A\otimes g$ ($\nabla
_{\lambda }g_{\mu \nu }=-2A_{\lambda }g_{\mu \nu }=-Q_{\lambda \mu
\nu }$), where  $A=A_{\mu }dx^{\mu }$ is a differential 1-form and
$Q$ is called the non-metricity \cite{AuGaSt84}. A change of
calibration of the metric induces a gauge transformation for $A$:
$g\rightarrow \exp \left( 2\lambda \right) g$, $A\rightarrow
A-d\lambda $. Therefore in terms of the torsion and non-metricity
the connection $\Gamma _{\mu \nu }^{\lambda }$ on $M$ can be
expressed as
\begin{equation}
\Gamma _{\mu \nu }^{\lambda }=\gamma _{\mu \nu }^{\lambda
}+T_{\text{ }\mu \nu }^{\lambda }+T_{\mu \nu }^{\text{ \ \ \
}\lambda }+T_{\nu \mu }^{\text{ \ \ \ }\lambda }+\frac{1}{2}\left(
Q_{\mu \nu }^{\text{ \ \ \ }\lambda }-Q_{\mu \text{ \ }\nu
}^{\text{ \ }\lambda }-Q_{\nu \text{ \ }\mu }^{\text{ \ }\lambda
}\right) ,
\end{equation}
where $\gamma _{\mu \nu }^{\lambda }$ is the Christoffel symbol
computed from the metric $g$ by using the general relativistic
prescription, $\gamma _{\alpha \beta }^{\mu }=g^{\mu \nu }\left(
\partial _{\alpha }g_{\nu \beta }+\partial _{\beta }g_{\nu \alpha
}-\partial _{\nu }g_{\alpha \beta }\right) $. Generally, we may
represent the connection as $\Gamma _{\mu \nu }^{\lambda }=\gamma
_{\mu \nu }^{\lambda }-K_{\nu \mu }^{\text{ \ \\}\lambda }$, where
$K_{\nu \mu }^{\text{ \ \\}\lambda }$, the contorsion tensor, is a
function of the torsion $T$ and of the non-metricity $Q$
\cite{He76}.

In the space-time $M$ the instantaneous state of a photon is given
by a four-momentum $p\in T_{x}\left( M\right) $ at an event $x\in
M$. The one-particle phase space $P_{\gamma }$ is a subset of the
tangent bundle given by \cite{Bi89}
\begin{equation}
P_{\gamma }:=\left\{ \left( x,p\right)\left|\right. x\in M,p\in
T_{x}\left( M\right) ,p^{2}=0\right\} .
\end{equation}

A state of a multiphoton system is described by a continuous,
non-negative distribution function $f\left( x,p\right) $, defined
on $P_{\gamma }$, and which gives the number $dN$ of
the particles of the system which cross a certain space-like volume $dV$ at $%
x$, and whose 4-momenta $p$ lie within a corresponding
three-surface element $d\vec{p}$ in the momentum space. The mean
value of $f$ gives the average number of occupied photon states
$\left( x,p\right) $. Macroscopic, observable quantities can be
defined as moments of $f$.

Let $\left\{ x^{\alpha }\right\} $, $\alpha =0,1,2,3$ be a local
coordinate system in $M$, defined in some open set $U\subset M$.
Then $\left\{ \partial /\partial x^{\alpha }\right\} $ is the
corresponding natural basis for tangent vectors. We express each
tangent vector $p$ in $U$ in terms of this basis as $p=p^{\alpha
}\frac{\partial }{\partial x^{\alpha }}$ and define a system of
local coordinates $\left\{ z^{A}\right\} $, $A=0,...,7$ in
$T_U(M)$ as $z^{\alpha }=x^{\alpha }$, $z^{\alpha +4}=p^{\alpha
}$. This defines a natural basis in the tangent space given by
$\left\{
\partial /\partial z^{A}\right\} =\left\{ \partial /\partial
x^{\alpha },\partial /\partial p^{\alpha }\right\} $. A vertical
vector field over $TM$ is given by $\pi =p^{\alpha }\partial
/\partial p^{\alpha }$. The geodesic flow field $\sigma $, which
can be constructed over the tangent bundle, is defined as $\sigma
=p^{\alpha }\partial /\partial x^{\alpha }-p^{\alpha }p^{\gamma
}\Gamma _{\alpha \gamma }^{\beta }\partial /\partial p^{\beta
}=p^{\alpha }D_{\alpha }$, where $D_{\alpha }=\partial /\partial
x^{\alpha }-p^{\gamma }\Gamma
_{\alpha \gamma }^{\beta }\partial /\partial p^{\beta }$. Physically, $%
\sigma $ describes the phase flow for a stream of particles whose
motion through space-time is geodesic.

Therefore the transport equation for the propagation of a photon
beam in a curved arbitrary non-Riemannian space-time is given by
\begin{equation}\label{distr}
\left( p^{\alpha }\frac{\partial }{\partial x^{\alpha }}-p^{\alpha
}p^{\beta }\Gamma _{\alpha \beta }^{i}\frac{\partial }{\partial
p^{i}}\right) f=0.
\end{equation}

By taking explicitly into account the decomposition of the
connection in an arbitrary non-Riemannian space-time with torsion
and non-metricity, Eq. (\ref{distr}) for the photon distribution
function takes the form
\begin{equation}
\left( p^{\alpha }\frac{\partial }{\partial x^{\alpha }}-p^{\alpha
}p^{\beta }\gamma _{\alpha \beta }^{i}\frac{\partial }{\partial
p^{i}}\right) f\left( x^{\mu },p^{\nu }\right) +p^{\alpha
}p^{\beta }K_{\alpha \beta }^{\text{ \ \\ }i}\frac{\partial
f\left( x^{\mu },p^{\nu }\right) }{\partial p^{i}}=0.
\end{equation}

To describe polarized or partially polarized radiation, the
distribution function has to be generalized to a distribution
tensor $f_{\alpha \beta
}\left( x,p\right) =\sum_{a,b=1}^{2}f_{ab}e_{\alpha }^{(a)}e_{\beta }^{(b)}$%
, where $e_{\alpha }^{(a)}$ ($a=1,2$) two space-like unit vectors
orthogonal to the direction of propagation. $f_{ab}$ is a
Hermitian and positive
matrix, satisfying the conditions Tr$\left( f_{ab}(x,p\right) =f(x,p)$, det$%
f_{ab}\left( x,p\right) \leq f(x,p)$ and $\bar{f}$ $_{ab}=f_{ba}$.
The matrix $f_{ab}$ can be parameterized by the Stokes parameters $\xi _{i}$,  $%
\xi _{i}:P_{\gamma }\rightarrow \left[ -1,1\right] $, $i=1,2,3$,
three measurable quantities, which describe polarization as
\cite{AcAn74}
\begin{equation}
f_{ab}=\frac{1}{2}f(x,p)\left(
\begin{array}{c}
1+\xi _{3}\text{ \ \ \ \ \ }\xi _{1}+i\xi _{2} \\
\xi _{1}+i\xi _{2}\text{ \ \ \ }1+\xi _{3}\text{\ \ \ }
\end{array}
\right) =\frac{1}{2}f(x,p)\left( 1+\xi \cdot \sigma \right) ,
\end{equation}
where $\sigma =$ $\left( \sigma _{1}\text{ }\sigma _{2}\text{
}\sigma
_{3}\right) $, with $\sigma _{i}$ the Pauli matrices,  and $\xi =\left( \xi _{1}%
\text{ }\xi _{2}\text{ }\xi _{3}\right) $.

In a space-like hypersurface element $\Omega $ of $P_{\gamma }$ at
$(x,p)$ the number of photons linearly polarized along the
$e^{(1)}$ and $e^{(2)}$
axes are $\frac{1}{2}\left( 1+\xi _{3}\right) f\Omega $ and $\frac{1}{2}%
\left( 1-\xi _{3}\right) f\Omega $. $\frac{1}{2}\left( 1+\xi
_{1}\right) f\Omega $ is the result of the measurement of the
linear polarization in a direction at $\pi /4$ to the $e^{(1)}$
axis. $P_{L}=\left( \xi _{1}^{2}+\xi
_{2}^{2}\right) ^{1/2}$ is the degree of linear polarization. The parameter $%
\xi _{2}$, the antisymmetric part of $f_{ab}$, represents the
degree of circular polarization. A measure for the right-hand and
left-hand circular polarization is $\left( 1+\xi _{2}\right) /2$
and $\left( 1-\xi _{2}\right) /2$, respectively. The polarization
degree is given by $P=\left( \sum_{i=1}^{3}\xi _{i}^{2}\right)
^{1/2}$. The polarization angle is defined as $\Psi =\tan \left(
\xi _{1}/\xi _{3}\right) /2$ \cite{Bi89}.

The transport equation satisfied by the distribution tensor
$f_{\mu \nu }\left( x,p\right) $, describing polarized radiation,
is \cite{Li66}, \cite{Bi89}
\begin{equation}
\left( p^{\alpha }\nabla _{\alpha }-p^{\alpha }p^{\beta }\Gamma
_{\alpha \beta }^{i}\frac{\partial }{\partial p^{i}}\right) f_{\mu
 \nu }(x,p)=0.
\end{equation}

In a general non-Riemannian space-time with non-vanishing torsion
and non-metricity the equation satisfied by the distribution
tensor takes the form
\begin{equation}
\left( p^{\alpha }\frac{\partial }{\partial x^{\alpha }}-p^{\alpha
}p^{\beta }\gamma _{\alpha \beta }^{i}\frac{\partial }{\partial
p^{i}}\right) f_{\mu \nu }+p^{\alpha }p^{\beta }K_{\alpha \beta
}^{\text{ \ \ \ }i}\frac{\partial f_{\mu \nu }}{\partial
p^{i}}-p^{\alpha }\gamma _{\mu \alpha }^{\lambda }f_{\lambda \nu
}-p^{\alpha }\gamma _{\nu \alpha }^{\lambda }f_{\mu \lambda
}+p^{\alpha }K_{\mu \alpha }^{\text{ \ }\lambda }f_{\lambda \nu
}+p^{\alpha }K_{\nu \alpha }^{\text{ \ }\lambda }f_{\mu \lambda
}=0.
\end{equation}

As in the Riemannian space-time we may introduce a
pseudo-orthonormal basis of vectors $e_{\alpha }=e_{\alpha
}^{a}\partial _{a}$ ( a tetrad), and the dual basis $\theta
^{\alpha }=e_{a}^{\alpha }dx^{a}$ of the one-forms. The components
$e_{\alpha }^{a}$ and their reciprocals $e_{a}^{\alpha }$satisfy
the relations $e_{\alpha }^{a}e_{b}^{\alpha }=\delta _{b}^{a}$ and $%
e_{\alpha }^{i}e_{i}^{\beta }=\delta _{\alpha }^{\beta }$,
respectively. The metric tensor can be expressed as
$g_{ab}=e_{a}^{\alpha }e_{b}^{\beta }\eta _{\alpha \beta }$, where
$\eta _{\alpha \beta }$ is the Minkowski metric
tensor \cite{EgGiHa80}. The object of anholonomity $\Omega _{ab}^{\text{ \ \ \ }%
c}=e_{a}^{\alpha }e_{b}^{\beta }\partial _{\lbrack i}e_{f]}^{\text{ \ \ }c}$%
, $\Omega _{abc}=$ $\Omega _{ab}^{\text{ \ \ \ }d}g_{dc}$ measures
the non-commutativity of the tetrad basis. The connection
expressed in these anholonomic coordinates is \cite{He76},
\cite{EgGiHa80}
\begin{equation}
\Gamma _{abc}=\Gamma _{ab}^{d}g_{dc}=-\Omega _{abc}+\Omega
_{bca}-\Omega _{cab}-K_{abc}.
\end{equation}

Any tangent vector $p$ at $x$ can be expressed as $p=p^{\alpha }e_{\alpha }$%
. If we set $e_{a}=e_{a}^{\alpha }e_{\alpha }$, then
$p^{a}=e_{\alpha }^{a}p^{\alpha }$ and $p^{\alpha }=e_{a}^{\alpha
}p^{a}$. In terms of the tetrad components the transport equations
can be expressed in the forms
\begin{equation}
p^{a}D_{a}f=0,p^{a}D_{a}f_{bc}-p^{a}\Gamma
_{ba}^{d}f_{dc}-p^{a}\Gamma _{ca}^{d}f_{bd}=0,
\end{equation}
where $D_{a}=\partial _{a}-\Gamma _{ac}^{b}p^{c}\partial /\partial
p^{b}$.

As an application of the tetrad formalism briefly described above
we consider the transformation from the coordinate basis
$e_{\alpha }=\partial _{\alpha }$ to the basis $b_{\alpha }$ of an
arbitrary coordinate system, obtained from
a Lorentz boost of the coordinate basis, so that $b_{\alpha }=\hat{L}%
_{\alpha }^{\beta }e_{\beta }$, $e_{\alpha }=L_{\alpha }^{\beta }b_{\beta }$%
, with the transformation matrices $\hat{L}_{\alpha }^{\beta }$ and $%
L_{\alpha }^{\beta }$ satisfying the condition $L_{\mu }^{\alpha }\hat{L}%
_{\alpha }^{\beta }=\delta _{\mu }^{\beta }$. The explicit form of
the matrix components is given by $\hat{L}_{0}^{b}=u^{b}$,
$\hat{L}_{\alpha }^{0}=\lambda a_{\alpha }u_{\alpha }$ and
$\hat{L}_{\alpha }^{\beta }=a_{\alpha }\left( \delta _{\alpha
}^{\beta }+ku^{\beta }u_{\alpha }\right) $, where $a_{\alpha
}=\sqrt{g^{\alpha \alpha }}$, $\lambda =\sqrt{-g^{00}}$ and
$k=\left( 1+u^{0}\sqrt{-g_{00}}\right) ^{-1}$. The components of
the four-momentum with respect to $b_{\alpha }$ \ are
$\hat{p}^{\alpha }=L_{\beta }^{\alpha }p^{\beta }$. The photon
four-momentum is described by its frequency, defined as $\nu
=-p\cdot u=-\hat{p}^{a}b_{a}\cdot u=\hat{p}^{0} $, and two angle
for the direction of propagation. Hence, the photon four-momentum
can be described in the basis $\left\{ b_{\alpha }\right\} $as
$\hat{p}^{\alpha }=\nu \left( 1,\hat{n}\right) $. To parameterize
the unit three-vector $\hat{n}$ we introduce spherical coordinates
in the momentum space so that
\begin{equation}
\hat{n}^{1}=\mu ,\hat{n}^{2}=\sqrt{1-\mu ^{2}}\cos \phi ,\hat{n}^{3}=\sqrt{%
1-\mu ^{2}}\sin \phi .
\end{equation}

By assuming that the observer places a two-dimensional screen
normal to the propagation direction of the photon beam in his rest
frame, we can perform a
further transformation of the basis so that in the new basis $d_{\alpha }=%
\hat{D}_{\alpha }^{\beta }b_{\beta }$ the spatial propagation
direction of the photon is now the $z$-axis. The transformation
matrix $\hat{D}_{\alpha }^{\beta }$ is given by \cite{Bi89a}
\begin{equation}
\hat{D}_{\alpha }^{\beta }=\left(
\begin{array}{cccc}
1 & 0 & 0 & 0 \\
0 & \sqrt{1-\mu ^{2}} & 0 & \mu  \\
0 & -\mu \cos \phi  & \sin \phi  & \sqrt{1-\mu ^{2}}\cos \phi  \\
0 & -\mu \sin \phi  & -\cos \phi  & \sqrt{1-\mu ^{2}}\sin \phi
\end{array}
\right) .
\end{equation}

The components of the four-momentum in the new basis $\left\{
d_{\alpha }\right\} $are $\bar{p}^{\alpha }=\nu \left(
1,0,0,1\right) $. With respect
to the initial basis $e_{\alpha }=\partial _{\alpha }$we have $d_{\alpha }=%
\hat{A}_{\alpha }^{\beta }e_{\beta }$, where $\hat{A}_{\alpha }^{\beta }=%
\hat{L}_{\mu }^{\beta }\hat{D}_{\alpha }^{\mu }$. The inverse
transformation is given by $e_{\alpha }=A_{\alpha }^{\beta
}d_{\beta }$. The components of
the four-momentum in the two basis are related by the transformations $\bar{p%
}^{\alpha }=A_{\beta }^{\alpha }p^{\beta }=D_{\beta }^{\alpha }\hat{p}%
^{\beta }$.

With respect to the basis $\left\{ d_{\alpha }\right\} $ the
transport equation can be written as
\begin{equation}
\left( \hat{A}_{\beta }^{\alpha }\bar{p}^{\beta }\frac{\partial
}{\partial x^{\alpha }}-\bar{p}^{\alpha }\bar{p}^{\beta
}\bar{\gamma}_{\alpha \beta
}^{i}\frac{\partial }{\partial \bar{p}^{i}}\right) \bar{f}_{ab}+\bar{p}%
^{\alpha }\bar{p}^{\beta }\bar{K}_{\alpha \beta }^{\text{ \ \ \ }i}\frac{%
\partial \bar{f}_{ab}}{\partial \bar{p}^{i}}-\bar{p}^{\alpha }\bar{\gamma}%
_{a\alpha }^{c}\bar{f}_{cb}-\bar{p}^{\alpha }\bar{\gamma}_{b\alpha }^{c}\bar{%
f}_{ac}+\bar{p}^{\alpha }\bar{K}_{a\alpha }^{c}\bar{f}_{cb}+\bar{p}^{\alpha }%
\bar{K}_{b\alpha }^{c}\bar{f}_{ac}=0,
\end{equation}
where $\bar{K}_{\alpha \beta }^{\text{ \ \ \ }i}$ are the
components of the contorsion tensor with respect to $\left\{
d_{\alpha }\right\} $. The new
connection coefficients $\bar{\gamma}_{\alpha \beta }^{\lambda }$ and $\bar{K%
}_{\alpha \beta }^{\text{ \ \ \ }\lambda }$can be calculated by
means of the transformations
\begin{equation}
\bar{\gamma}_{\alpha \beta }^{\lambda }=A_{\mu }^{\lambda
}\hat{A}_{\beta }^{\eta }\left( \hat{A}_{\alpha }^{\nu }\gamma
_{\eta \nu }^{\mu }+\partial _{\eta }\hat{A}_{\alpha }^{\mu
}\right) ,
\end{equation}
and
\begin{equation}
\bar{K}_{\alpha \beta }^{\text{ \ \ \ }\lambda }=A_{\mu }^{\lambda }\hat{A}%
_{\beta }^{\eta }\left( \hat{A}_{\alpha }^{\nu }\bar{K}_{\eta \nu
}^{\text{ \ \ \ }\mu }+\partial _{\eta }\hat{A}_{\alpha }^{\mu
}\right) ,
\end{equation}
respectively.

By denoting
\begin{equation}
D=\frac{1}{\nu }\left( \hat{A}_{\beta }^{\alpha }\bar{p}^{\beta }\frac{%
\partial }{\partial x^{\alpha }}-\bar{p}^{\alpha }\bar{p}^{\beta }\bar{\gamma%
}_{\alpha \beta }^{i}\frac{\partial }{\partial \bar{p}^{i}}+\bar{p}^{\alpha }%
\bar{p}^{\beta }\bar{K}_{\alpha \beta }^{\text{ \ \ \ }i}\frac{\partial \bar{%
f}_{ab}}{\partial \bar{p}^{i}}\right) ,
\end{equation}
it follows that in a non-Riemannian space-time the radiation
distribution function and the Stokes parameters satisfy the
equations
\begin{equation}\label{pol1}
Df=0,D\xi _{1}-2\left(
\bar{\gamma}_{20}^{1}+\bar{\gamma}_{23}^{2}\right) \xi
_{3}+2\left( \bar{K}_{20}^{1}+\bar{K}_{23}^{2}\right) \xi _{3}=0,
\end{equation}
\begin{equation}
D\xi _{2}=0,D\xi _{3}+2\left( \bar{\gamma}_{20}^{1}+\bar{\gamma}%
_{23}^{2}\right) \xi _{1}-2\left(
\bar{K}_{20}^{1}+\bar{K}_{23}^{2}\right) \xi _{1}=0.
\end{equation}

The evolution equations describing the degrees of the linear polarization $%
P_{L}$ and the total polarization $P$ are
\begin{equation}\label{pol2}
DP_{L}=0,DP=0.
\end{equation}

In the limit of the zero contorsion $K_{\alpha \beta }^{\text{ \ \ \ }%
i}\rightarrow 0$, corresponding to the transition to the
Riemannian geometry, from Eqs. (\ref{pol1})-(\ref{pol2}) we
recover the transport equations given in \cite{Bi89,Bi89a}.

\section{Radiation transfer in isotropic and homogeneous Weyl-Cartan space-times}

In order to analyze the influence of the non-Riemannian background
on the propagation properties of the electromagnetic radiation we
adopt the simplifying assumption of an isotropic, homogeneous and
flat Friedmann-Robertson-Walker type geometry, that is, we assume
that the metric of the Universe is given by
\begin{equation}
ds^{2}=dt^{2}-a^{2}(t)\left( dx^{2}+dy^{2}+dz^{2}\right) .
\end{equation}

Generally in the comoving reference frame the torsion tensor is
determined by two functions of time $T(t)$ and $\bar{T}\left(
t\right) $, so that $T_{0\alpha \beta }=\left( T(t)/l\right)
\delta _{\alpha \beta }$ and $T_{\alpha \beta \gamma }=\left(
\bar{T}\left( t\right) /l\right) \varepsilon _{\alpha \beta \gamma
}$, where $l$ is a coupling constant and $\varepsilon _{\alpha
\beta \gamma }$ is the totally antisymmetric tensor in three
dimensions \cite{ItHe03}. By supposing that the theory is
invariant under space inversions we have $\bar{T}\left( t\right)
\equiv 0$. Therefore the only non-vanishing components of the torsion are $%
T(t)=T_{\text{ }10}^{1}=T_{\text{ }20}^{2}=T_{\text{ }30}^{3}$.
The non-metricity torsion is defined by three functions of time $Q_{i}(t),i=1,2,3$%
, so that, in an anholonomic basis we have $\bar{Q}_{110}=\bar{Q}_{220}=\bar{Q%
}_{330}=Q_{1}(t)$, $\bar{Q}_{000}=Q_{2}(t)$ and $\bar{Q}_{011}=\bar{Q}_{022}=%
\bar{Q}_{033}=Q_{3}(t)$ \cite{MiGa98}.

By denoting $K(t)=-\left[T(t)+Q_{3}(t)\right]$, the transport
equation of the total polarization degree $P$ of the gamma-rays in
a non-Riemannian space-time is given by
\begin{equation}\label{part}
\left[ \frac{\partial }{\partial t}+\frac{\mu }{a}\frac{\partial }{\partial x%
}+\frac{\sqrt{1-\mu ^{2}}\cos \phi }{a}\frac{\partial }{\partial y}+\frac{%
\sqrt{1-\mu ^{2}}\sin \phi }{a}\frac{\partial }{\partial z}-\nu \left( \frac{%
\dot{a}}{a}-K(t)\right) \frac{\partial }{\partial \nu }\right]
P=0.
\end{equation}

We assume that the photon is emitted at a point $Q=\left(
x_{0},y_{0},z_{0}\right) $ with frequency $\nu _{0}$ in the direction $%
\left( \mu _{0},\phi _{0}\right) $ and is observed at the point
$O=\left( x,y,z\right) $ with the corresponding four-momentum
quantities $\left( \nu ,\mu ,\phi \right) $.

Eq. (\ref{part}) represents a first order partial differential
equations with the characteristics given by
\begin{equation}
dt=\frac{adx}{\mu }=\frac{ady}{\sqrt{1-\mu ^{2}}\cos \phi }=\frac{adz}{\sqrt{%
1-\mu ^{2}}\sin \phi }=-\frac{d\nu }{\nu \left( \frac{\dot{a}}{a}%
+K(t)\right) },
\end{equation}
which can be immediately integrated to give
\begin{equation}
\sqrt{1-\mu ^{2}}\cos \phi x-\mu y=\sqrt{1-\mu _{0}^{2}}\cos \phi
_{0}x_{0}-\mu _{0}y_{0}=\text{constant},
\end{equation}
\begin{equation}
\sqrt{1-\mu ^{2}}\sin \phi x-\mu z=\sqrt{1-\mu _{0}^{2}}\sin \phi
_{0}x_{0}-\mu _{0}z_{0}={\rm constant},
\end{equation}
\begin{equation}
\sin \phi y-\cos \phi z=\sin \phi _{0}y_{0}-\cos \phi
_{0}z_{0}={\rm constant},
\end{equation}
\begin{equation}\label{red}
\ln \left( a\nu \right) +\int K(t)dt={\rm constant},
\end{equation}
\begin{equation}
x-\mu \int \frac{dt}{a}={\rm constant},y-\sqrt{1-\mu ^{2}}\cos \phi \int \frac{dt%
}{a}={\rm constant},z-\sqrt{1-\mu ^{2}}\sin \phi \int
\frac{dt}{a}={\rm constant}.
\end{equation}

Hence the general solution of Eq. (\ref{part}) is given by
\begin{eqnarray}\label{sol}
P &=&P\left(\sqrt{1-\mu ^{2}}\cos \phi x-\mu y,\sqrt{1-\mu
^{2}}\sin \phi x\right.-\mu z,\sin \phi y-\cos \phi z,x-\mu \int
\frac{dt}{a}, \nonumber\\
 &&\left.y-\sqrt{1-\mu ^{2}}\cos \phi
\int \frac{dt}{a},z-\sqrt{1-\mu ^{2}}\sin \phi \int
\frac{dt}{a},\ln \left( a\nu \right) +\int K(t)dt\right).  \
\end{eqnarray}

For $K(t)\equiv 0$, Eq. (\ref{red}) gives the usual redshift
relation $a\nu =a_{0}\nu _{0}$ and Eq. (\ref{sol}) describes the
propagation of radiation in the Riemannian space-time of general
relativity.

In the following we denote by $P_{0}$ the value of the
polarization corresponding to the propagation of the
electromagnetic radiation in a curved Riemannian space-time,
$P_{0}=\left.P\right|_{K(t)\equiv 0}$. We also make the
simplifying assumption that generally the dependence of the
polarization on
the term $\ln \left( a\nu \right) +\int K(t)dt$ is linear, so that
\begin{equation}
P=P_0\left[ 1+\frac{1}{\ln \left( a\nu \right)}\int K(t)dt/)
\right].
\end{equation}

In order to obtain a quantitative characterization of the
non-Riemannian effects on the radiation propagation we introduce a
parameter $\delta $, describing the variation of the polarization
due to the propagation effects, and defined as
\begin{equation}
\delta =\frac{P-P_{0}}{P_{0}}\text{.}
\end{equation}

In terms of the contorsion tensor, by taking into account that at
the receiver $a=a_{rec}$ and by denoting the observed frequency of
the radiation by $\nu _{rec}$, we obtain
\begin{equation}
\delta =\frac{1}{\ln \left( a_{rec}\nu _{rec}\right) }\int K(t)dt.
\end{equation}

The integration over $t$ can be converted to integration over the
redshift $z$, by using the equality $dt=(dt/da)(da/dz)dz=-d_{H}\left( z\right) dz/(1+z)$%
, where $d_{H}(z)=\left( \dot{a}/a\right) ^{-1}=H^{-1}(z)$, where
$H$ is the Hubble function. We assume that generally the Hubble
function can be written as
\begin{equation}
H\left( z\right) =H_{0}\sqrt{\Omega _{m,0}\left( 1+z\right)
^{3}+\Omega _{\Lambda }+\Omega _{K,0}f(z)},
\end{equation}
where $H_{0}=3.24\times 10^{-18}h$ s$^{-1}$, with $0.5<h<1$,
$\Omega _{m,0}\approx 0.3$ is the present day matter density
parameter, $\Omega _{\Lambda }\approx 0.7$ is the dark energy
density parameter and $\Omega _{K,0}$ is the density parameter
formally associated with the non-Riemannian effects. $f\left(
z\right) $ describes the time variation of the contorsion. $\Omega
_{K,0}$ and $f(z)$ are strongly model dependent quantities and
generally they depend on the assumed functional form for the
contorsion $K(t)$. Therefore in terms of the redshift, the
parameter $\delta $ can be expressed as
\begin{equation}
\delta (z)=-\frac{t_{H}}{\ln \left( a_{rec}\nu _{rec}\right) }\int_{z}^{0}\frac{%
K(z)dz}{\left( 1+z\right) \sqrt{\Omega _{m,0}\left( 1+z\right)
^{3}+\Omega _{\Lambda }+\Omega _{K,0}f(z)}},
\end{equation}
where $t_{H}=1/H_{0}=3.09\times 10^{17}h^{-1}$ s is the Hubble
time.

\section{Constraining the torsion and the non-metricity with Gamma Ray
Bursts}

As an application of the formalism developed in the previous
Section we consider the possible effects of a non-Riemannian
structure of the space-time on the propagation of the gamma-ray
bursts. Gamma-ray bursts (GRBs) are cosmic gamma ray emissions
with typical fluxes of the order of $10^{-5}$ to $5\times
10^{-4}$erg cm$^{-2}$ with the rise
time as low as $10^{-4}$ s and the duration of bursts from $10^{-2}$ to $%
10^{3}$ s. The distribution of the bursts is isotropic and they
are believed to have a cosmological origin, recent observations
suggesting that GRBs might originate at extra-galactic distances
\cite{ChLu01}. The large inferred distances imply isotropic energy
losses as large as $3\times 10^{53} $ erg for GRB 971214 and
$3.4\times 10^{54}$ erg for GRB 990123 \cite {Me02}.

The widely accepted interpretation of the phenomenology of $\gamma
$-ray bursts is that the observable effects are due to the
dissipation of the kinetic energy of a relativistically expanding
fireball, whose primal cause is not yet known \cite{ZhMe03}.

The proposed models for the energy source involve merger of binary
neutron stars, capture of neutron stars by black holes,
differentially rotating neutron stars  or neutron star-quark star
conversion etc. (\cite{ZhMe03} and references therein). However,
the most popular model involves the violent formation of an
approximately one solar mass black hole, surrounded by a similarly
massive debris torus. The formation of the black hole and debris
torus may take place through the coalescence of a compact binary
or the collapse of a quickly rotating massive stellar core
\cite{Me02}. There are still many open problems concerning GRBs,
from which the most important is the problem of the source of the
large energy emission during the bursts.

The recent observations of the polarization of the GRB's are
considered to be of fundamental importance for the understanding
of the nature and properties of these phenomena. The polarization
of the afterglows is well established. Typically, the polarization
of the afterglows is observed to be at the $1-3\%$ level, with
constant or smoothly variable level and position angle when
associated with a relatively smooth light curve. The usual
explanation for the polarized radiation of afterglows is the
synchrotron emission from the fireball when some kind of asymmetry
is present. There are mainly two kinds of asymmetry considered.
One class includes the models assuming that ordered magnetic
fields play a crucial role. The magnetic fields can be either
locally ordered, which corresponds to the magnetic domain model
\cite{GrWa99}, or even entirely aligned within the ejecta, which
is magnetized by the central engine \cite{GrKo03}. In addition,
small regions in which the magnetic field has some degree of order
could be amplified by scintillation \cite{MeLo99}, or by
gravitational micro-lensing \cite{LoPe98}. A different class of
models postulates that the fireball is collimated. In this case
the observer likely sees the fireball off-axis, as the probability
of being exactly on-axis is vanishingly small. When the line of
sight makes an angle with the collimation axis, the asymmetry and
hence a net polarization may arise. The magnetic fields,
compressed in the plane normal to the motion, are postulated to be
distributed in a random but anisotropic manner \cite{MeLo99},
\cite{Wi99}.

A strong gamma-ray polarization may indicate a strongly magnetized
central engine, either in pure Poynting-flux-dominated form
\cite{Ly03} or in conventional hydrodynamical form, but with a
globally organized magnetic field configuration \cite{Wa03}.
Models involving inverse Compton scattering with offset beaming
angles can also give rise to large degrees of polarization in
gamma-rays \cite{Sh95}. There are also indications that the
polarization degree and position angle may evolve significantly
with time for optical transients \cite{Gr03}. Two-component jet
models have also been proposed to explain the observed
polarization of the afterglows \cite{Wu05}.

Hence still there are no definite physical models to consistently
predict or explain the observed polarization of the gamma ray
bursts.

Since a non-Riemannian geometrical structure of the space-time
manifold cannot be excluded {\it a priori}, the possible effect of
a non-zero contorsion must also be taken into account in the
description of the propagation of the gamma ray bursts. In the
following we investigate the effect of different choices of the
torsion and non-metricity, corresponding to different physical and
cosmological models, on the parameter $\delta =(P-P_0)/P_0$,
describing the deviations in the polarization of the gamma-ray
bursts due to the propagation effects in a Weyl-Cartan geometry.

\subsection{Models with constant contorsion}

The simplest case that could arise in the analysis of the
modifications of the polarization of the gamma-ray bursts due to
propagation effects is the one corresponding to a constant
contorsion, with $K(z)=K_{A}=$constant. Moreover, we ignore the
contributions of the torsion and non-metricity to
the density parameter, by taking $\Omega _{K,0}=0$. Therefore we obtain for $%
\delta $ the simple expression
\begin{equation}\label{s1}
\delta (z)=-\frac{t_{H}K_{A}}{\ln \left( \nu _{rec}\right) }\int_{z}^{0}%
\frac{dz}{\left( 1+z\right) \sqrt{\Omega _{m,0}\left( 1+z\right)
^{3}+\Omega _{\Lambda }}}.
\end{equation}

For this model the variation of $\delta $ as a function of $z$ is
represented, for different numerical values of the constant
contorsion $K_0$, in Fig. 1.

\vspace{0.2in}
\begin{figure}[h]
\includegraphics{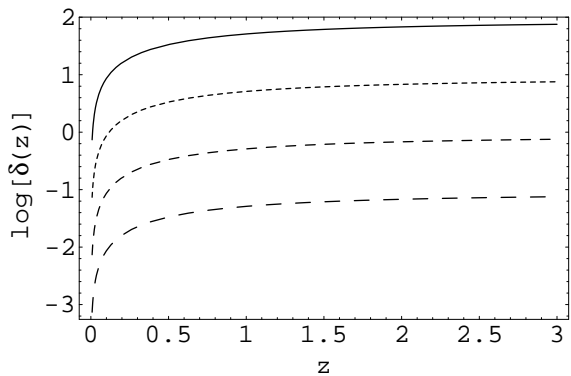}
\caption{Variation, as a function of the redshift $z$, of the
parameter $\delta $ (in a logarithmic scale) for gamma-ray bursts
propagating in a Weyl-Cartan geometry with constant contorsion
tensor $K_A$, for different values of $K_A$:
$K_A=10^{-14}$s$^{-1}$ (solid curve), $K_A=10^{-15}$s$^{-1}$
(dotted curve), $K_A=10^{-16}$s$^{-1}$ (dashed curve) and
$K_A=10^{-17}$s$^{-1}$ (long dashed curve). For the cosmological
parameters we have adopted the values $a_{rec}=1$, $\Omega
_{m,0}=0.3$ and $\Omega _{\Lambda }=0.7$. The frequency at the
detector of the gamma ray bursts is assumed to be $\nu
_{rec}=3\times 10^{14}$s$^{-1}$, corresponding to the optical
afterglow emission.} \label{FIG1}
\end{figure}

On the other hand the observational knowledge of $\delta $ would
allow to impose some constraints on the value of the contorsion
tensor, with
\begin{equation}
K_{A}\leq \frac{H_{0}\ln \left( a_{rec}\nu _{rec}\right)
}{F(z)}\delta ,
\end{equation}
where we denoted $F(z)=\int_{0}^{z}dz/\left( 1+z\right)
\sqrt{\Omega _{m,0}\left( 1+z\right) ^{3}+\Omega _{\Lambda }}$.

\subsection{The Weyssenhoff spin fluid case}

In the framework of the Einstein-Cartan theory, the intrinsic
angular momentum (the spin) of a particle is introduced via a
generalization of the structure of the space-time, by assuming
that the affine connection is nonsymmetric. The torsion
contributes to the energy-momentum of a spin fluid which has the
form
\begin{equation}
 T_{\mu \nu }^{{\rm eff}}=\left( \rho +p-2s^{2}\right) u_{\mu
}u_{\nu }-\left( p-s^{2}\right) g_{\mu \nu },
\end{equation}
where $\rho $ and $p$ are the energy density and the pressure,
respectively, and $s^{2}=s_{\mu \nu }s^{\mu \nu }$ is the spin.
This energy-momentum tensor corresponds to a perfect fluid with
spin, called the Weyssenhoff fluid model \cite{He76}. The
macroscopic spin tensor may be expressed in terms of the spin
density tensor $s_{\mu \nu }$ and the four-velocity of the fluid
as $\tau _{\mu \nu }^{\lambda }=u^{\lambda }s_{\mu \nu }$. For a
pressureless fluid ($p=0$) we have $\rho =mn$ and $s=\hbar n/2$,
respectively, where $n$ is the number of particles with mass $m$
per unit volume and $s$ is the spin density.  The effect of the
spin is dynamically equivalent to introducing into the model some
additional non-interacting spin fluid with the energy density
$\rho _s=\rho _{0s}/a^6$ \cite{SzKr04}. Although the contribution
of the spin to the dense matter appears to be negligible small, on
a large scale it can produces a "centrifugal force" which is able
to prevent the occurrence of singularities in cosmology
\cite{He76}.

By taking into account that for a Weyssenhoff fluid dominated
universe the contorsion is proportional to the spin,  the
parameter $\delta $ becomes
\begin{equation}\label{s2}
\delta (z)=-\frac{t_{H}K_{B}}{\ln \left( \nu _{rec}\right) }\int_{z}^{0}%
\frac{\left( 1+z\right) ^{2}dz}{\sqrt{\Omega _{m,0}\left(
1+z\right) ^{3}+\Omega _{s,0}\left( 1+z\right) ^{6}+\Omega
_{\Lambda }}},
\end{equation}
where $\Omega _{s,0}=\rho _{s,0}/3H_{0}^{2}$ is the density
parameter of the spin fluid, with $\rho _{s,0}=\hbar ^{2}n(0)/16$.
$S_{0}$ can be related to the mass $m$ of the particles forming
the spinning fluid by means of the relation $K_{B}=\rho
_{s,0}/\sqrt{2}m$.

The variation of the parameter $\delta $ for a Weyssenhoff fluid
filled universe is represented, for different values of the
cosmological parameters, in Fig. 2.

\vspace{0.2in}
\begin{figure}[h]
\includegraphics{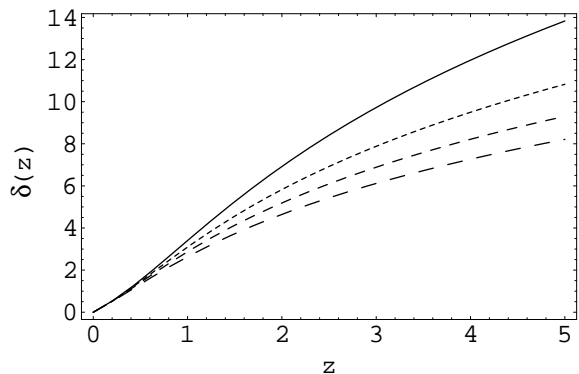}
\caption{The parameter $\delta $ as a function of the redshift $z$
for gamma-ray bursts propagating in a Weyssenhoff spinning fluid
filled universe, with the torsion tensor proportional to the spin
density, for different values of the cosmological parameters:
$\Omega _{m,0}=0.25$, $\Omega _{\Lambda}=0.7$ and $\Omega
_{s,0}=0.05$ (solid curve), $\Omega _{m,0}=0.20$, $\Omega
_{\Lambda}=0.7$ and $\Omega _{s,0}=0.1$ (dotted curve), $\Omega
_{m,0}=0.15$, $\Omega _{\Lambda}=0.7$ and $\Omega _{s,0}=0.15$
(short dashed curve) and $\Omega _{m,0}=0.20$, $\Omega
_{\Lambda}=0.6$ and $\Omega _{s,0}=0.20$ (long dashed curve). The
frequency at the detector of the gamma ray bursts is assumed to be
$\nu _{rec}=3\times 10^{14}$s$^{-1}$, corresponding to the optical
afterglow emission. For the constant $K_B$ we have adopted the
value $K_B=2.5\times 10^{-16}{\rm s}^{-1}$} \label{FIG2}
\end{figure}

The study of the polarization of the gamma ray bursts can impose
some constraints on $s$. Once the functional forms of the
polarization of the gamma ray bursts at the source and observer
are known, from Eq. (\ref{s2}) one can evaluate the basic physical
parameters characterizing the Weyssenhoff fluid and obtain some
restrictions on the spin density of the particles.

\subsection{Models with modified double duality ansatz}

An extensive study of the spatially homogeneous and
$SO(3)$-isotropic cosmologies in the framework of the $10$
parameter Lagrangian of the Poincar\'e gauge theory was performed
in \cite{GoMu84}. By evaluating the action in the tangent spaces
induced by the isotropy group it follows that the torsion tensor
has two non-vanishing components $h(t)$ and $f(t)$. By excluding
the parity violating terms in the Lagrangian the field equations
are invariant under the discrete transformation $f\rightarrow -f$ and $%
s\rightarrow -s$, where $s$ is the spin scalar. This symmetry
rules out classical spin distributions like the Weyssenhoff fluid
we have previously discussed. It is difficult to find a physical
interpretation of a spin tensor within this model and therefore
the case of a vanishing spin was considered in detail.

A simple class of solutions of the Einstein-Cartan field equations
can be obtained by assuming that the axial torsion is
non-vanishing and the modified double duality ansatz $\Phi =\Phi _{0}=$%
constant and $\Psi =0$ is satisfied, where $\Phi =M+N=R/6$ and
$\Psi =fH+F=R_{\alpha \beta \gamma \delta }\varepsilon ^{\alpha
\beta \gamma
\delta }$, with $H=h+\dot{a}/a$, $M=\dot{H}+\left( \dot{a}/a\right) H$, $%
N=H^{2}+k/a^{2}-f^{2}/4$ and $F=\left[ \dot{f}+\left( \dot{a}/a\right) f%
\right] $. By further imposing the constraints $H=0$ and $k=0$, it
follows that the only non-vanishing component of the torsion is
$h=-\dot{a}/a$, and the field equations reduce to $\left(
\dot{a}/a\right) ^{2}=\left( 4\pi G/9c_{4}\right) \rho
-c_{0}/6c_{4}$, $c_{0},c_{4}=$constants, and the
conservation law for the energy and pressure. In this model the function $%
K(t)$, describing the effect of the torsion on the propagation of
the gamma ray bursts is given by $K(t)=-K_{C}\ln a(t)$, where
$K_{C}$ is an integration constant. In terms of the redshift we
have $K(z)=K_{C}\ln \left( 1+z\right) $. The constants
$-c_{0}/6c_{4}$ can be interpreted as an effective cosmological
constant $\Lambda _{eff}$, generated by the presence of the
torsion. Therefore in this model the parameter $\delta \left(
z\right) $ is given by
\begin{equation}
\delta (z)=-\frac{t_{H}K_{C}}{\ln \left( \nu _{rec}\right) }\int_{0}^{z}%
\frac{\ln \left( 1+z\right) dz}{\left( 1+z\right) \sqrt{\Omega
_{m,0}\left( 1+z\right) ^{3}+\Omega _{\Lambda }}},
\end{equation}
where $\Omega _{\Lambda }=\Lambda _{eff}/\rho _c$ is the mass
density parameter associated to the torsion-generated cosmological
constant.

The variation of $\delta $ as a function of the redshift is
presented in Fig. 3.

\vspace{0.2in}
\begin{figure}[h]
\includegraphics{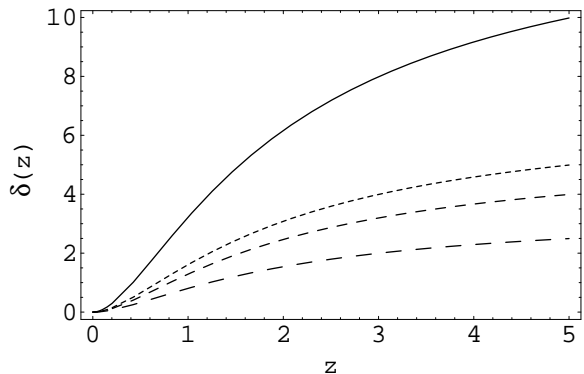}
\caption{The parameter $\delta $ as a function of the redshift $z$
for gamma-ray bursts propagating in a Riemann-Cartan space-time,
with vanishing spin density and the modified double duality ansatz
satisfied, for different values of the constant $K_C$:
$K_C=2\times 10^{-15}$s$^{-1}$ (solid curve), $K_C=
10^{-15}$s$^{-1}$ (dotted curve), $K_C=8\times 10^{-16}$s$^{-1}$
(dashed curve) and $K_C=5\times 10^{-16}$s$^{-1}$ (long dashed
curve). For the cosmological parameters we have adopted the values
$a_{rec}=1$, $\Omega _{m,0}=0.3$ and $\Omega _{\Lambda }=0.7$. The
frequency at the detector of the gamma ray bursts is assumed to be
$\nu _{rec}=3\times 10^{14}$s$^{-1}$, corresponding to the optical
afterglow emission.} \label{FIG3}
\end{figure}

\section{Discussions and final remarks}

In the present paper we have generalized the radiation propagation
equations for polarized photons in curved space-times from
Riemannian geometries to Weyl-Cartan type geometries, with torsion
and non-metricity. The non-Riemannian background enters in the
equations via the contorsion tensor and, for particles propagating
on cosmological distances, can induce a significant and
observationally detectable change in the photon polarization
state. Of course, to fully evaluate this effect one should know
the behavior of the contorsion in different classes of generalized
gravitation theories with torsion. For the Poincar\'e gauge theory
and for vanishing non-metricity spatially homogeneous and
$SO(3)$-isotropic exact solutions for a $10$-parameter Lagrangian
have been studied in \cite{GoMu84}, while the effect of the
non-metricity on the cosmological evolution has been considered in
\cite{MiGa98}. All these solutions are strongly model-dependent
and generally have a complicated mathematical form.

On the other hand the study of the polarization of the prompt
cosmological gamma ray bursts and of their afterglows offers, at
least in principle, the possibility of observationally testing
Poincar\'e type gauge theories.

Assuming, as an extreme and perhaps unrealistic case, that the
initial gamma ray emission is totally unpolarized, as predicted by
some models of the gamma ray burst emission (see
\cite{ChLu01}-\cite{Pi04} and references therein), the detected
polarization amount could be mainly due to the propagation effects
of the photons travelling on cosmological distances in a
non-Riemannian geometrical background. The presence of the
contorsion in the radiation transport equations in curved
space-times could generate a supplementary polarization of the
photon beam. If the polarization of the gamma ray bursts is indeed
due to the coupling between photons and geometry, then the
observed degree of the polarization must be redshift-dependent,
since photons travelling on longer distances will be more
affected. The smallness of the observed effect even for high
redshift sources ($P=1.7\pm0.2\%$) also seems to suggest mainly a
propagation effect.

A second important effect, which follows from the consideration of
the non-Riemannian structure of the space-time is related to the
prediction of the frequency dependence of the polarization of
gamma ray bursts. Since the dependence of the frequency is
logarithmic, the total polarization degree of the gamma ray
component should be smaller by a factor of around $2$ than the
polarization degree of the optical afterglow emission. Together
with the  cosmological redshift dependence, the frequency
dependence of the polarization degree of gamma ray bursts provides
a clear observational signature of the possible presence of
non-Riemannian, Weyl-Cartan type geometrical features in our
Universe.

On the other hand, if the initial polarization degree of the gamma
ray bursts could be exactly predicted by some photon emission
models, then the comparison of the polarization of the gamma ray
bursts at the observer and emitter could impose some strong
constraints on the deviations of the geometry of the Universe with
respect to the Riemannian background. The torsional effects
generated by the spin fluid modify the background geometry and
affects the propagation of the photons. The study of the
polarization of the gamma ray bursts could also impose some limits
on the spin density and overall rotation of the Universe.

However, the practical implementation of an observational program
aiming at detecting torsion and non-metricity from the
polarization of the gamma ray bursts could be extremely difficult.
From the present observational point of view, the polarization
measurements of the gamma ray bursts and the interpretation of the
data are still controversial. There is no general agreement on the
detected value of the polarization degree \cite{CoBo03},
\cite{RuFo04}, \cite{Wi04}, and, since at the moment the data
quality is insufficient to constrain the polarization degree in
most of the cases \cite{Wi04}, a major improvement in the
observational techniques is required. On the other hand, there are
many astrophysical mechanisms which can induce an initial
polarization of the photons from the cosmological gamma ray
bursts, which makes the task of distinguishing between the
different involved physical processes extremely difficult. The
propagation of the photons takes place on cosmological distances
in a cosmic environment whose properties also present many
uncertainties. However, when a larger amount of high precision
spectropolarimetric data from gamma ray bursts will be available,
the possibility of testing the foundations of general relativity
and other more general gravitational theories could become a
reality.

\section*{Acknowledgments}

We would like to thank Prof. Y. F. Huang for very useful
suggestions and help. This work is supported by a grant of the
government of the Hong Kong SAR.

\end{document}